\documentclass[a4paper]{jpconf}
\usepackage{graphicx}
\usepackage{rotating}
\usepackage{subfigure}

\begin{document}

\title{Measurement of charm and bottom production in p+p collisions at $\sqrt{s}$ = 200 GeV at RHIC-PHENIX} 

\author{Yuhei Morino for the PHENIX Collaboration }

\address{Center for Nuclear Study, Graduate School of Science, University of Tokyo,\\
 7-3-1, Hongo, Bunkyo, Tokyo 113-0033, Japan\\}

\ead{y.morino@cns.s.u-tokyo.ac.jp}

\begin{abstract}
RHIC-PHENIX has observed a large suppression pattern and azimuthal 
anisotropy of non-photonic electron at mid-rapidity~($\mid\eta\mid<0.35$)
in Au+Au collisions at $\sqrt{s_{NN}} = 200$ GeV.
To understand these results and the interaction of heavy quarks in the 
hot and dense medium, experimental determination of production ratio 
of charm over bottom is one of the most important topics, 
since the behavior of bottom may differ from charm in the medium.
We measured the ratio of charm over bottom  and total cross section of bottom 
via partial reconstruction of D$^0$$\rightarrow$e$^+$  K$^-$  $\nu_e$ decay 
in p+p collisions at $\sqrt{s} = 200$ GeV.
Total cross sections of charm and bottom were also measured via di-electron continuum 
in p+p collisions at $\sqrt{s} = 200$ GeV.  
\end{abstract}

\section{Introduction}
Heavy quarks~(charm and bottom) are good probes to study the hot and dense
matter created Au+Au collisions at RHIC, since heavy quarks can only be produced in
initial collisions.
PHENIX experiment has measured heavy quarks via measurement of non-photonic
electrons with $0.3 < p_{\mathrm{T}} < 9.0$~GeV/$c$ at
mid-rapidity~($\mid\eta\mid<0.35$).
A large suppression pattern and azimuthal anisotropy 
of non-photonic electrons have been observed in Au+Au collisions 
at $\sqrt{s_{NN}} = 200$ GeV~\cite{bib1,bib12}.
To understand these results and the interaction of heavy quarks in the 
medium, a ratio of the number of electrons from charm over that from bottom 
is one of the most important parameters.
A fixed-order-plus-next-to-leading-log~(FONLL) perturbative QCD calculation predicts 
electrons from bottom will dominate these from
charm above 4~GeV/$c$~\cite{bib2}. However, there is a large uncertainty about 
cross section of heavy flavor at even FONLL.
Thus, experimental determination of this ratio is a key issue.

\section{Partial reconstruction of D$^0$}
PHENIX has measured the ratio of the number of electron from charm over that from bottom via 
 partial reconstruction of D$^0$$\rightarrow$e$^+$  K$^-$  $\nu_e$ decay in p+p 
collisions at $\sqrt{s}=200$~GeV.
Unlike charge sign pairs of electrons with  $2.0 < p_{\mathrm{T}}<7.0$~GeV/$c$ 
and hadrons with $0.4 < p_{\mathrm{T}}<5.0$~GeV/$c$ were 
reconstructed as partial reconstruction of 
D$^0$$\rightarrow$e$^+$  K$^-$ $\nu_e$ decay.
Kaon identification for hadrons was not performed for high statistics.
Like charge sign pairs of electrons and hadrons were used for 
background subtraction. Since charge asymmetry of electrons and hadrons 
was only produced by weak decay, combinatorial background and contributions from
photonic electrons were canceled out completely. Most of contribution from jet was 
also canceled out, since jet was basically charge independent. 
\begin{figure}[thb]
  \centering
  \subfigure[]{
    \includegraphics[angle=0,width=5.5cm]{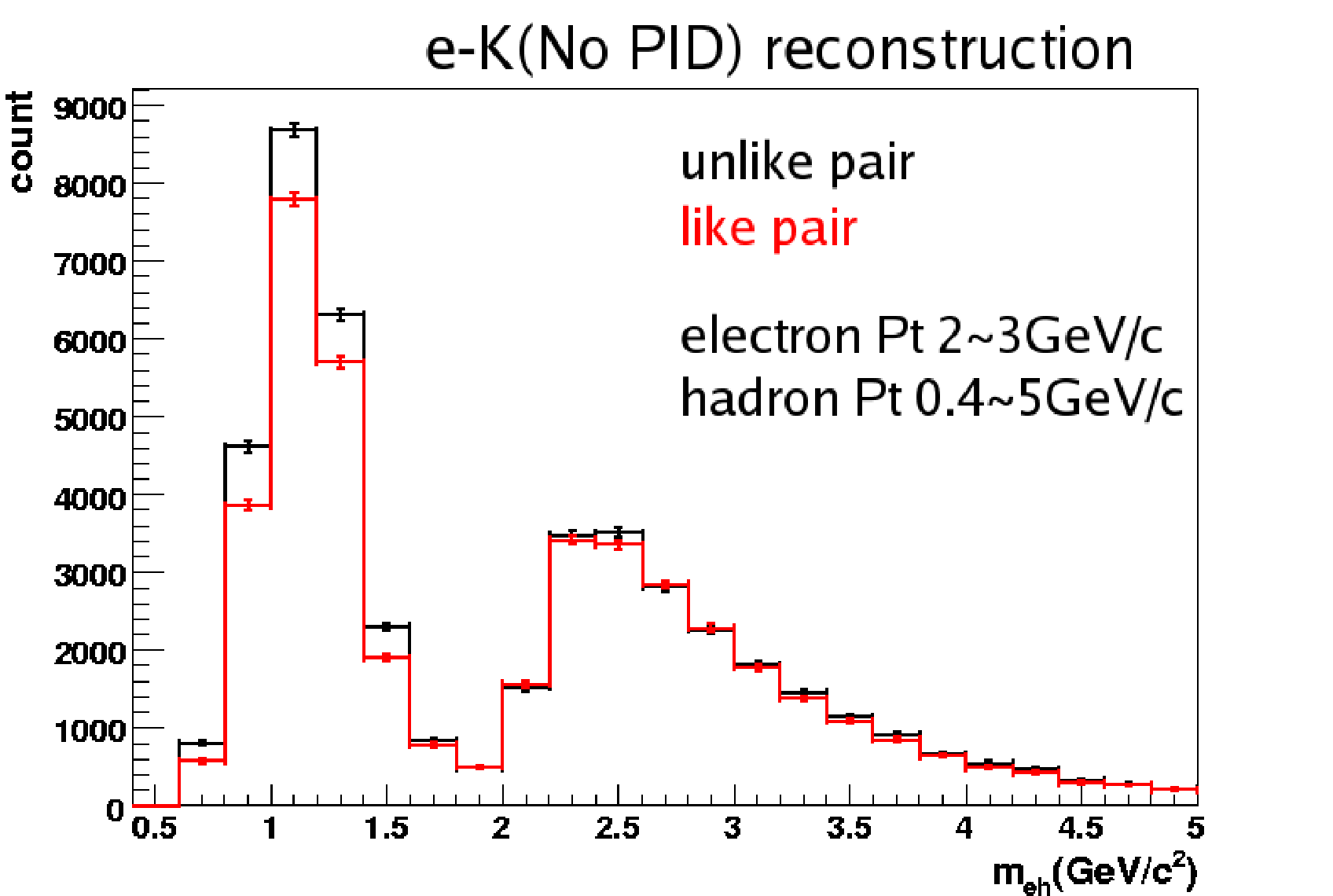}
    \label{fig1a}
  }
  \hspace{0.7cm}
  \subfigure[]{
    \includegraphics[angle=0,width=5.5cm]{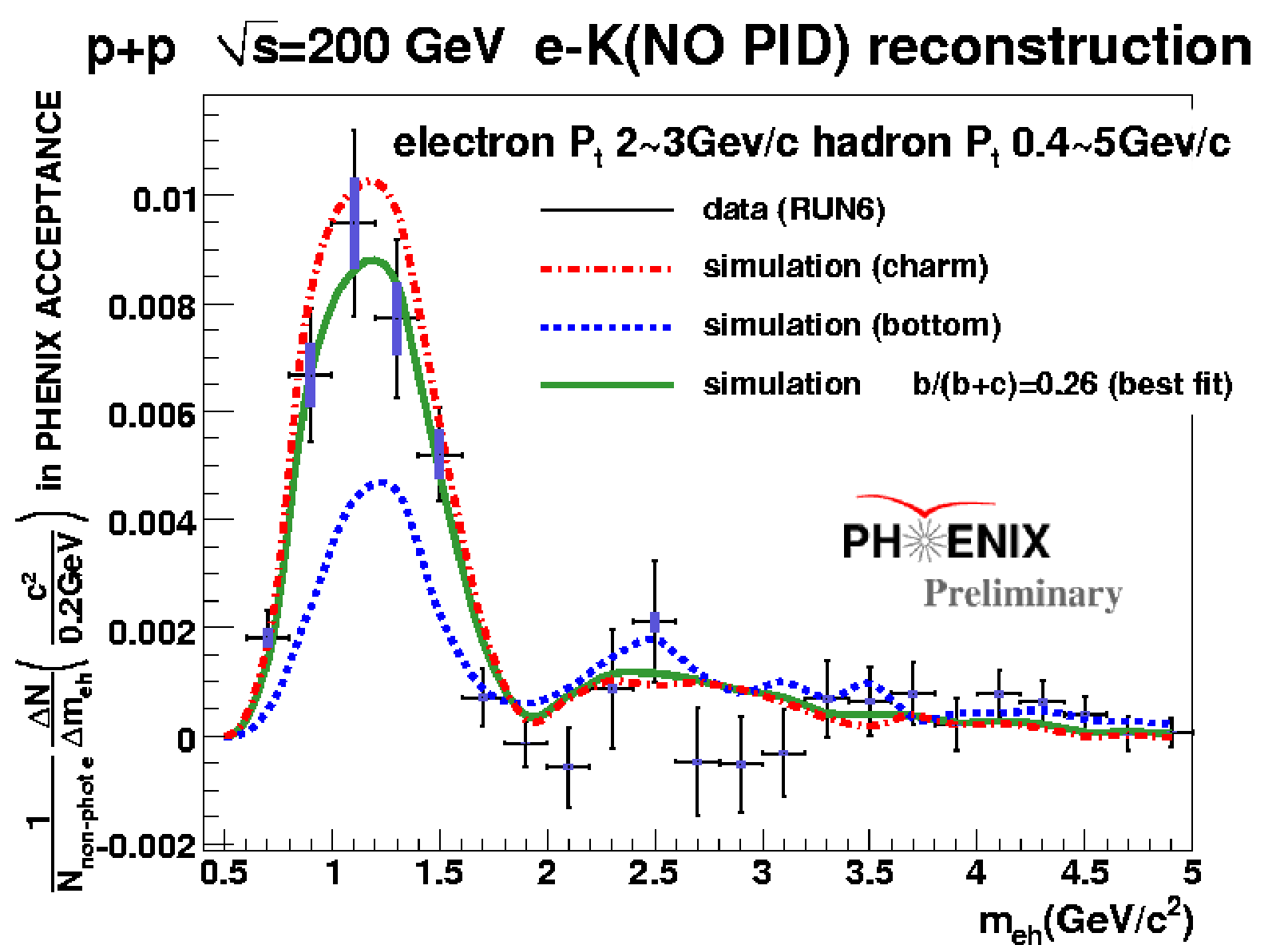}
    \label{fig1b}
  }
  \caption{(a)Raw invariant mass distribution of 
    electrons and hadrons. (b) Reconstruction signals with the mass distributions 
    of charm and bottom from PYTHIA and the mass distribution 
    when charm and bottom components are mixed at the obtained ratio.
  }
  \label{fig1}
\end{figure}

Figure~\ref{fig1a} shows raw invariant mass distribution of
electrons and hadrons.
N$_{tag}$ was defined as the number of unlike sign pair~(N$_{unlike}$) minus
the number of like sign pair~(N$_{like}$) in the mass range from 0.4 to 1.9~GeV/$c^2$.
Tagging efficiency~($\epsilon_{data}$) was defined as follows, by using the number 
of non-photonic electrons~(N$_{e}$).
\[ 
\epsilon_{data} \equiv \frac{N_{tag}}{N_e} 
= \frac{N_{c\rightarrow tag}+N_{b\rightarrow tag}}
{N_{c\rightarrow e}+N_{b\rightarrow e}}
\]
Here, c~(b) means charm~(bottom). 
The ratio of the number of electrons from charm over that from bottom was determined as follows.
\[
 \epsilon_c \equiv \frac{N_{c\rightarrow tag}}{N_{c\rightarrow e}}, \,\,
 \epsilon_b \equiv \frac{N_{b\rightarrow tag}}{N_{b\rightarrow e}},
 \]
 \[ 
 \frac{b\rightarrow e}{c\rightarrow e+b\rightarrow e} = 
 \frac{\epsilon_c -\epsilon_{data}}{\epsilon_c-\epsilon_b}
\]
Here, $\epsilon_{c~(b)}$ is a tagging efficiency in the case of charm~(bottom) production.
$\epsilon_{c~(b)}$ was calculated from PYTHIA and EvtGen simulation~\cite{bib3,bib4}.\\
Extracted signals~(N$_{tag}$) contain not only signals from daughters 
of semi-leptonic decay of D$^0$, but also these from daughters 
of semi-leptonic decay of other heavy flavored hadons and jet contributions which
were not canceled completely.
These contributions were taken into account by PYTHIA.
Since dominant component of $\epsilon_{c~(b)}$ was daughters 
of semi-leptonic decay of heavy flavored hadrons and cross sections of heavy flavors 
in PYTHIA did not have any contribution for $\epsilon_{c~(b)}$,
uncertainty of $\epsilon_{c~(b)}$ from PYTHIA was reduced at this method.
\begin{figure}[thb]
  \centering
  \subfigure[]{
    \includegraphics[angle=0,width=6.5cm]{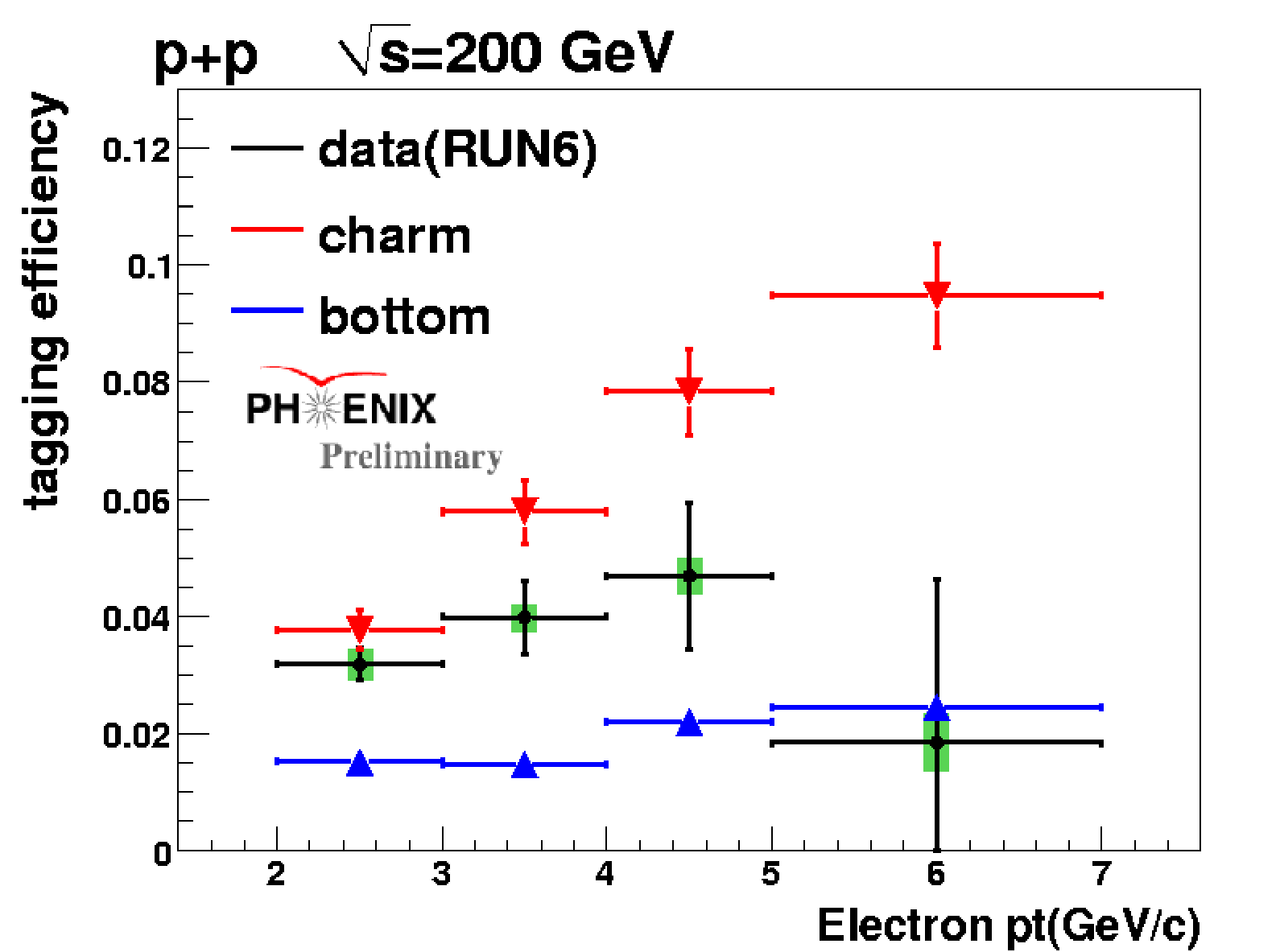}
    \label{fig2a}
  }
  \hspace{0.7cm}
  \subfigure[]{
    \includegraphics[angle=0,width=7.5cm]{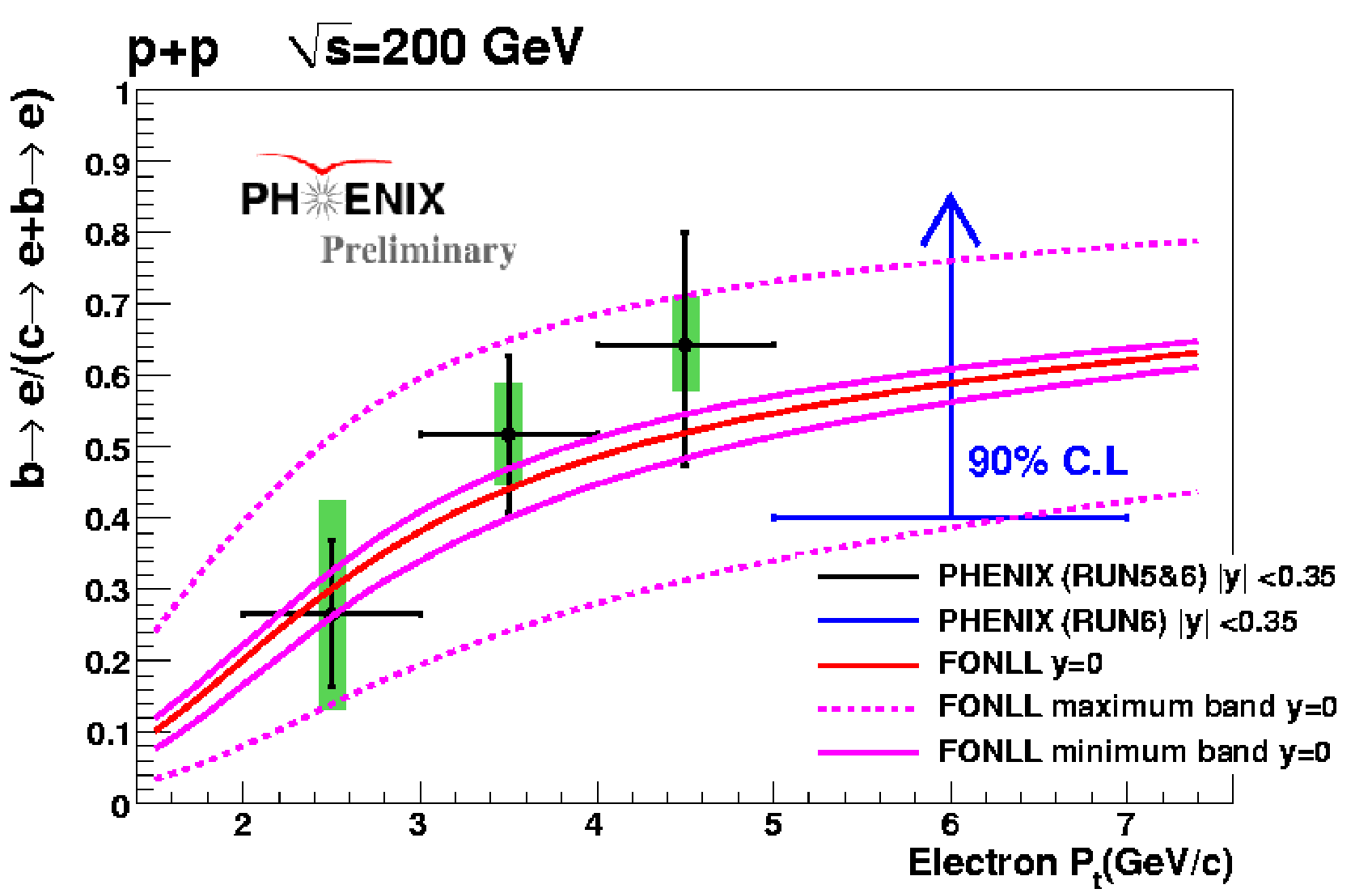}
    \label{fig2b}
  }
  \caption{(a)$\epsilon_{data}$, $\epsilon_{c}$ and $\epsilon_{b}$ as a 
function of electron $p_{\mathrm{T}}$.(b)Ratios of the number of electrons from charm over that from bottom  
as a function of electron $p_{\mathrm{T}}$ with FONLL prediction }
  \label{fig2}
\end{figure}
Figure~\ref{fig1b} shows reconstruction signals with the mass distributions 
of charm and bottom from PYTHIA and the mass distribution 
when charm and bottom components are mixed at the obtained ratio.
Figure~\ref{fig2a} shows $\epsilon_{data}$, $\epsilon_{c}$ and $\epsilon_{b}$ as a 
function of electron $p_{\mathrm{T}}$.
Figure~\ref{fig2b} shows the obtained ratios of the number of electrons from charm over that from bottom 
as a function of electron $p_{\mathrm{T}}$ with FONLL prediction.
The measured ratio is consistent with FONLL predictions.
Bottom contribution becomes dominant above electron $p_{\mathrm{T}} \sim 3$~GeV/$c$.\\

\begin{figure}[thb]
  \centering
  \includegraphics[angle=0,width=7.5cm]{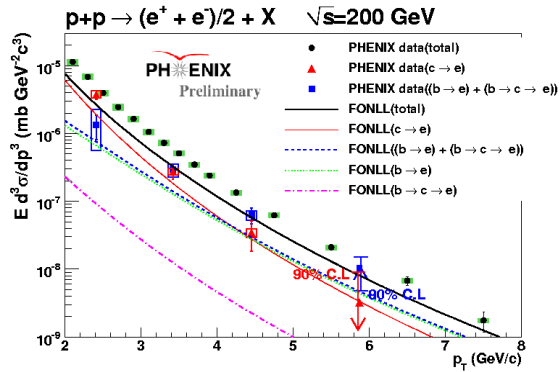}
  \caption{Electron spectra from charm and bottom with FONLL
    predictions.}
  \label{fig3}
\end{figure}
Electron spectra from charm and bottom were obtained from the ratio 
of the number of electrons from charm over that from bottom and the spectrum of non-photonic 
electrons~\cite{bib12}.
Figure~\ref{fig3} shows obtained electron spectra from charm and bottom with FONLL
predictions.
The ratio of bottom cross section at PHENIX over predicted value at 
FONLL is about 2, which is almost same as
HERA-B and CDF~\cite{bib5,bib6}.
Electron spectrum from bottom was extrapolated to $p_{\mathrm{T}}=0$ by using the spectrum
shape predicted by FONLL~\cite{bib7}.
The bottom cross section at mid-rapidity was $d\sigma_{b\bar{b}}/dy_{\mid y=0} =
 1.34 \pm 0.38(stat) ^{+0.74}_{-0.64}(sys) \mu b$,
by using a b $\rightarrow$ e total branching ratio of $10 \pm 1 \%$
Total bottom cross section was obtained as $\sigma_{b\bar{b}}=
   4.61 \pm 1.31(stat) ^{+2.57}_{-2.22}(sys) \mu b$, 
by using rapidity distribution from NLO pQCD calculation~\cite{bib8}.

\section{Di-electron continuum}
\begin{figure}[thb]
  \centering
  \subfigure[]{
    \includegraphics[angle=0,width=7cm]{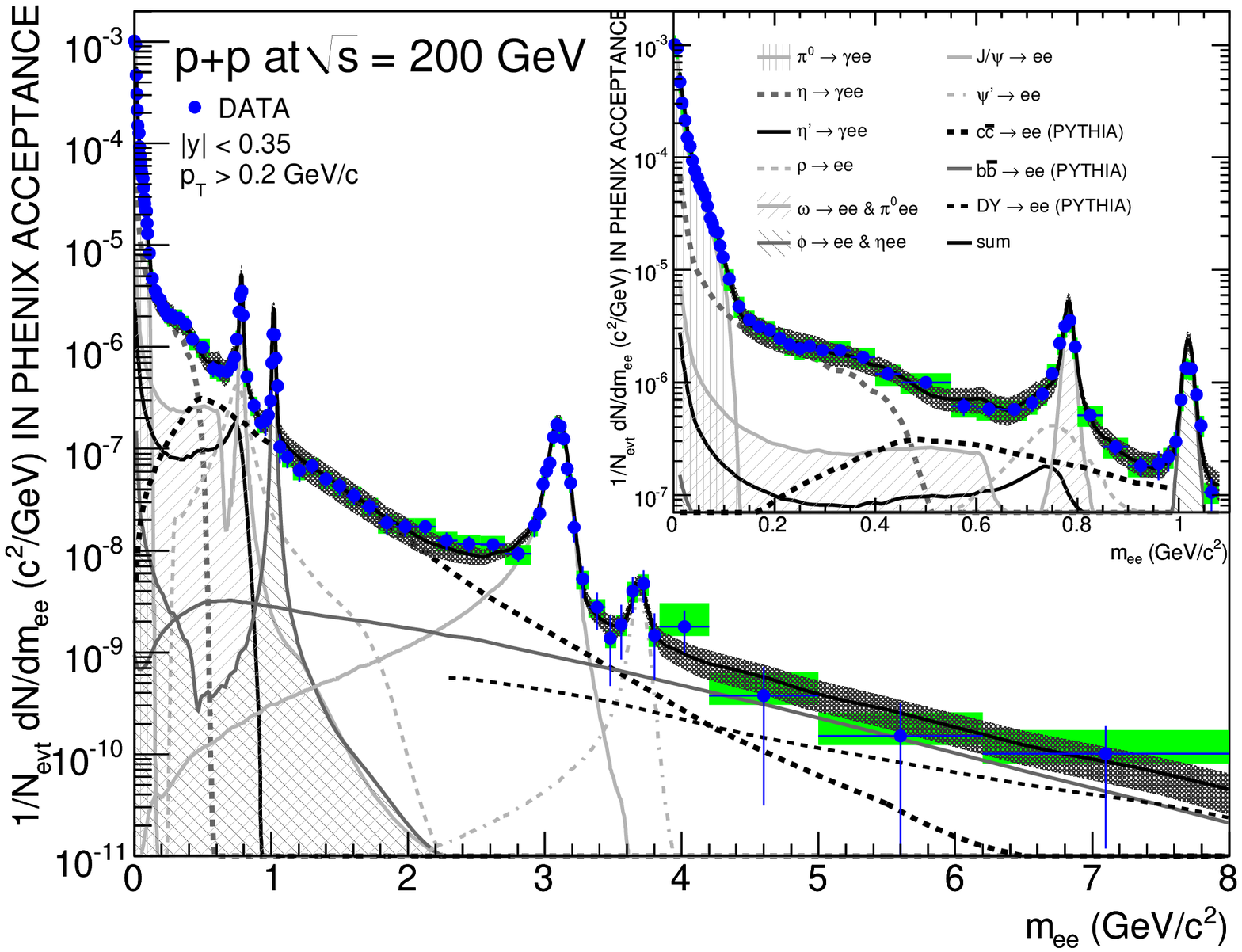}
    \label{fig4a}
  }
  \hspace{0.7cm}
  \subfigure[]{
    \includegraphics[angle=0,width=7cm]{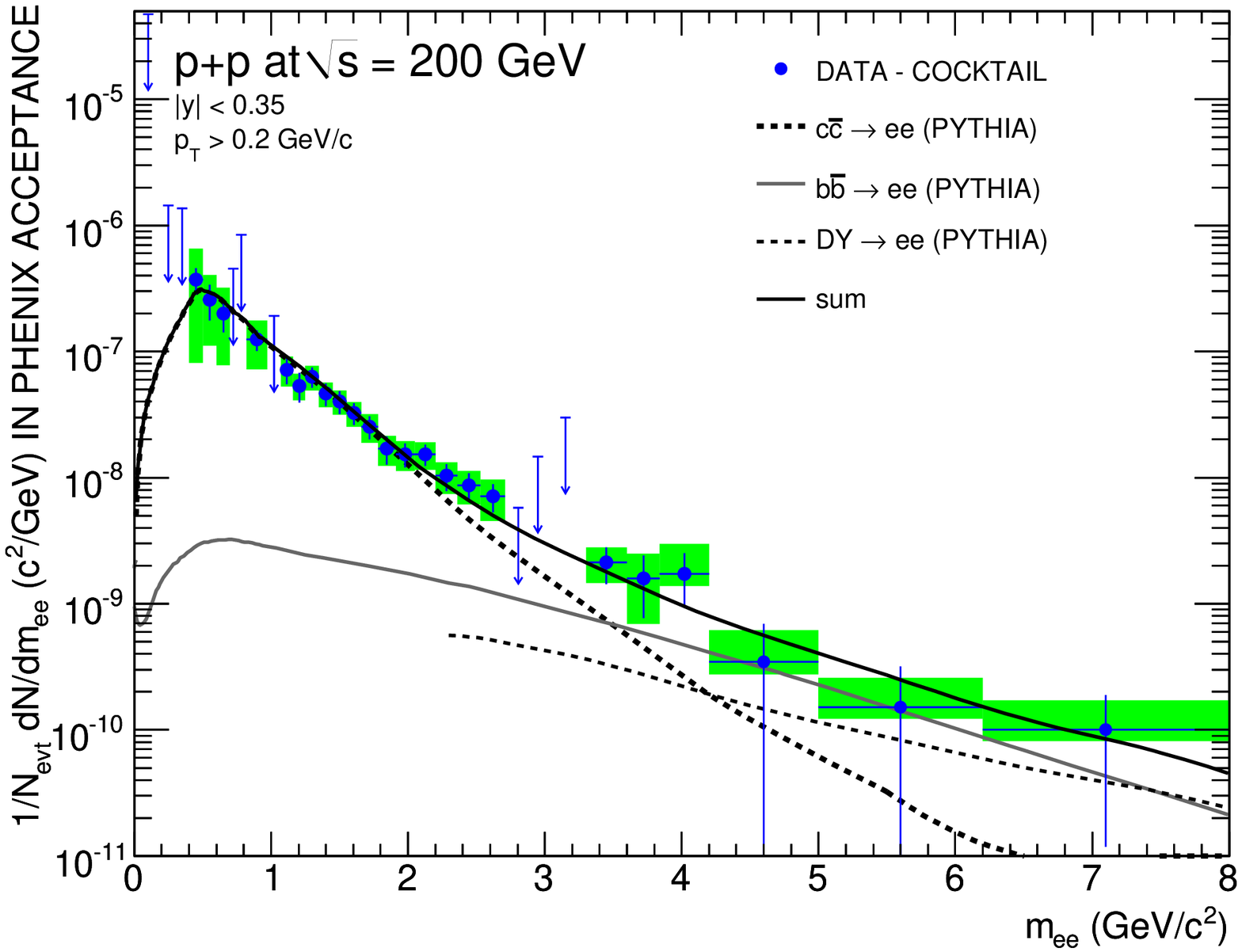}
    \label{fig4b}
  }
  \caption{(a)The $e^-e^+$ pair yield per p+p collision 
    in PHENIX acceptance with a cocktail of known sources.
  (b)The $e^-e^+$ pair yield remaining after subtraction of 
    the cocktail.}
  \label{fig4}
\end{figure}

PHENIX has measured the electron-positron pair mass spectrum in
p+p and Au+Au collisions at $\sqrt{s}=200$~GeV~\cite{bib9,bib10}.
Figure~\ref{fig4a} shows measured $e^-e^+$ pair yield per p+p collision 
in PHENIX acceptance with a cocktail of known sources. 
The cocktail accounts for nearly the continuum in the mass region below $\sim$
1~GeV/$c^2$. Except for the vector meson peaks, the $e^-e^+$ pair in the mass range
above 1.1~GeV/$c^2$ is dominated by heavy quarks correlated through flavor conservation.\\
Figure~\ref{fig4b} shows the $e^-e^+$ pair yield remaining after subtraction of 
the cocktail. The remaining components are $c\bar{c}$, $b\bar{b}$ and Drell-Yan process.
To extract signals from $c\bar{c}$, the $e^-e^+$ pair yield in the range from 1.1 to 2.5
GeV/$c^2$ was integrated.
Integrated yield was extrapolated to zero $e^-e^+$ pair mass by using PYTHIA simulation
and converted cross section of charm.
Contributions from $b\bar{b}$ and Drell-Yan process were estimated and subtracted. 
Total cross section of charm was obtained as  $\sigma_{c\bar{c}}=
   544 \pm 39(stat) \pm 142(sys) \pm 200(model) \mu b$, 
by using rapidity distribution from NLO pQCD calculation~\cite{bib8}.
This result is compatible with PHENIX previous result of non-photonic electron 
which gave $\sigma_{c\bar{c}}=
   567 \pm 57(stat) \pm 224(sys) \mu b$~\cite{bib12}.\\
Alternative method was used to extract signals from $b\bar{b}$.
The $e^-e^+$ pair distribution after subtraction of Drell-Yan was fitted
by the $e^-e^+$ pair distributions from charm and bottom which were produced by PYTHIA.
The obtained total cross section of bottom is $\sigma_{b\bar{b}}=
   3.9 \pm 2.5(stat) ^{+3}_{-2}(sys) \mu b$.
This result is also compatible with the result of non-photonic electron in 
Section~2.

\section{Summary and Outlook}
The ratio of the number of electron from charm over that from bottom was measured 
via partial reconstruction of D$^0$$\rightarrow$e$^+$  K$^-$  $\nu_e$decay 
in p+p collisions at $\sqrt{s} = 200$ GeV.
Bottom contribution becomes dominant above electron $p_{\mathrm{T}} \sim 3.5$~GeV/$c$.
This suggests not only charm but also bottom may lose large energy in the hot
and dense medium.
Total cross section of bottom was obtained from the ratio and non-photonic electron
spectrum.\\
Total cross sections of charm and bottom were also measured via di-electron continuum.
The result from di-electron continuum are consistent with that from non-photonic electron spectra.\\
Direct reconstruction of D$^0$ meson has been tried at PHENIX.
Clear peaks has been observed in D$^0$ $\rightarrow$ 
K$^-$ $\pi^+$ $\pi^0$ and D$^0$ $\rightarrow$ K$^-$ $\pi^+$ channels in
p+p collisions at $\sqrt{s} = 200$ GeV.
Cross check of charm production will be performed by direct reconstruction.


\section*{References}
\begin{thebibliography}{11}  
  
\bibitem{bib1} A.~Adare {\it et al.}(PHENIX Collaboration) {\it Phys. Rev. Lett.} {\bf 98}, 172301 (2007) 
\bibitem{bib12}A.~Adare {\it et al.} (PHENIX Collaboration){\it Phys. Rev. Lett.} {\bf 97}, 252002 (2006) 
\bibitem{bib2} M.~Cacciari {\it et al.} {\it Phys. Rev. Lett.} {\bf 95}, 122001 (2005) 
\bibitem{bib3} We used PYTHIA 6.403 with parameters as in R.Field.
\bibitem{bib4} We used EvtGen v00-11-07 with default parameters.
\bibitem{bib5} D.~Acosta {\it et al.} {\it Phys. Rev. Lett} {\bf 91}, 241804 (2003) 
\bibitem{bib6} S.~Chekanov {\it et al.} {\it Eur. Phys. J} {\bf C50}, 299 (2007) 
\bibitem{bib7} M.~Cacciari, private communication.
\bibitem{bib8} M.~L.~Mangano, P.~Nason and G.Ridolfi {\it Nucl. Phys} {\bf B405}, 507 (1993)
\bibitem{bib9} A.~Adare {\it et al.}(PHENIX Collaboration) arXiv:0706.3034 
\bibitem{bib10}A.~Adare {\it et al.}(PHENIX Collaboration) arXiv:0802.0050 
\end{thebibliography}
\end{document}